\documentclass[twocolumn]{aastex62}
\usepackage{color,xcolor}

\graphicspath{{./}{figures/}}

\received{June 4, 2019}
\revised{July 17, 2019}
\accepted{for publication in ApJ: July 24, 2019}

\shorttitle{Ram-pressure stripping in A1758N}
\shortauthors{Ebeling \& Kalita}

\begin{document}

\title{Jellyfish: Ram-pressure stripping as a diagnostic tool in studies of cluster collisions\footnote{Data presented herein were obtained at the W.M.\ Keck Observatory, which is operated as a scientific partnership among the California Institute of Technology, the University of California, and the National Aeronautics and Space Administration. The observatory was made possible by the generous financial support of the W.M.\ Keck Foundation.}\thanks{Based on observations made with the NASA/ESA Hubble Space Telescope, obtained at the Space Telescope Science Institute, which is operated by the Association of Universities for Research in Astronomy, Inc., under NASA contract NAS 5-26555. These observations are associated with programs GO-14096.}}

\correspondingauthor{Harald Ebeling}
\email{ebeling@ifa.hawaii.edu}

\author{ Harald Ebeling \& Boris S.\ Kalita}
\affil{Institute for Astronomy\\
University of Hawaii\\
2680 Woodlawn Dr\\
Honolulu, HI 96822, USA}

\begin{abstract}
Prompted by the discovery of A1758N\_JFG1, a spectacular case of ram-pressure stripping (RPS) in the galaxy cluster A1758N, we investigate the properties of other galaxies suspected to undergo RPS in this equal-mass, post-collision merger. Exploiting constraints derived from Hubble Space Telescope images and Keck longslit spectroscopy, our finding of apparent debris trails and dramatically enhanced star formation rates in an additional seven RPS candidates support the hypothesis that RPS, and hence rapid galaxy evolution in high-density environments, is intricately linked to cluster collisions. Unexpectedly, we find the vast majority of RPS candidates in A1758N to be moving toward us, and in a shared direction as projected on the plane of the sky. We hypothesize that this directional bias is the result of two successive events: (1) the quenching, during and after the first core passage, of star formation in galaxies with an approximately isotropic velocity distribution within the central region of the merger, and (2) RPS events triggered in late-type galaxies falling into the merging system along a filament, possibly enhanced by a shock front expanding into the outskirts of the south-eastern subcluster. Since this explanation implies that the merger axis of A1758N must be significantly inclined with respect to the plane of the sky, our findings open the possibility of RPS events becoming important diagnostic tools to constrain the geometry of cluster collisions that, due to the orientation of the merger axis, lack the classic observational signatures of face-on mergers.
\end{abstract}


\section{Introduction} \label{sec:intro}

By providing a physical mechanism to remove both molecular and atomic gas from spiral galaxies moving through the diffuse gas filling the potential wells of massive clusters of galaxies, ram-pressure stripping \citep[hereafter RPS;][]{gunn72} has been recognized to play an important (and possibly dominant) role in the transformation of late-type into early-type galaxies. Supported by numerical simulations \citep[e.g.,][]{abadi99,vollmer01,roediger05,mccarthy08}, observational studies of RPS in nearby clusters have yielded extensive insight into the dynamics and efficiency of RPS \citep[e.g.,][]{kenney04,abramson11,fumagalli14} and, more recently, have begun to explore related topics, such as the significance of cluster mergers \citep[e.g.,][]{stroe15,mcpartland16,deshev17} and the interplay of RPS and nuclear activity in galaxies \citep{pog17,george19}. 

\begin{figure}
    \centering
    \includegraphics[width=0.45\textwidth]{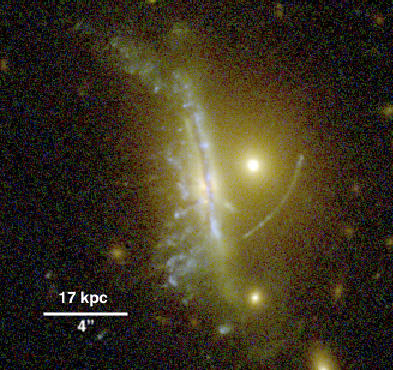}
    \caption{A1758N\_JFG1, a newly discovered spectacular example of ram-pressure stripping in the massive cluster merger A1758N ($z=0.279$) as viewed by HST/ACS. The unmistakable debris trails and edge-on view of the galactic disk allow an unambiguous determination of this ``jellyfish" galaxy's direction of motion relative to the intra-cluster medium. We refer to \citet{kalita19} for an in-depth review of the properties of A1758N\_JFG1.}  
    \label{fig:slit-a1758n-jfg1}
\end{figure}

We here continue our observational investigation of galaxy evolution in the dense environment provided by massive galaxy clusters, focusing on the characterization of ram-pressure stripping events at $z>0.2$, i.e., at redshifts beyond which the Universe is large enough to contain a significant number of truly massive clusters \citep[see also][]{cortese07,ebeling14}. The analysis of RPS events and their link to the dynamics and history of cluster mergers described in this paper was triggered by our discovery of a spectacular case of RPS in the merging double cluster A1758N, shown in Fig.~\ref{fig:slit-a1758n-jfg1} and discussed in detail in \citet{kalita19}.

Throughout this paper we adopt the concordance $\Lambda$CDM cosmology, characterised by  $\Omega_{m}=0.3$, $\Omega_{\Lambda}=0.7$, and $H_{0}=70$ km s$^{-1}$ Mpc$^{-1}$. All images are oriented such that north is up and east is to the left.

\section{Abell 1758}
\label{sec:target}
Abell 1758 \citep{abell58} is a rich cluster of galaxies at $z=0.28$ found to consist of two components, A1758S and A1758N, separated by about 8 arcmin on the sky. Both components are in turn merging systems that are well studied from X-ray to radio wavelengths \citep[e.g.,][]{david04,durret11,botteon18}. Although A1758S and A1758N are bound to merge eventually, no evidence of physical interaction is observed at their current separation of approximately 2 Mpc (in projection). We here focus exclusively on A1758N, an active merger of two similarly massive clusters.

\section{Observational Data}
\label{sec:obs}

We here briefly summarize the data used (or referred to) in this work.

\subsection{HST imaging}
\label{sec:hst-obs}

\begin{figure*}
    \centering
    \includegraphics[width=\textwidth]{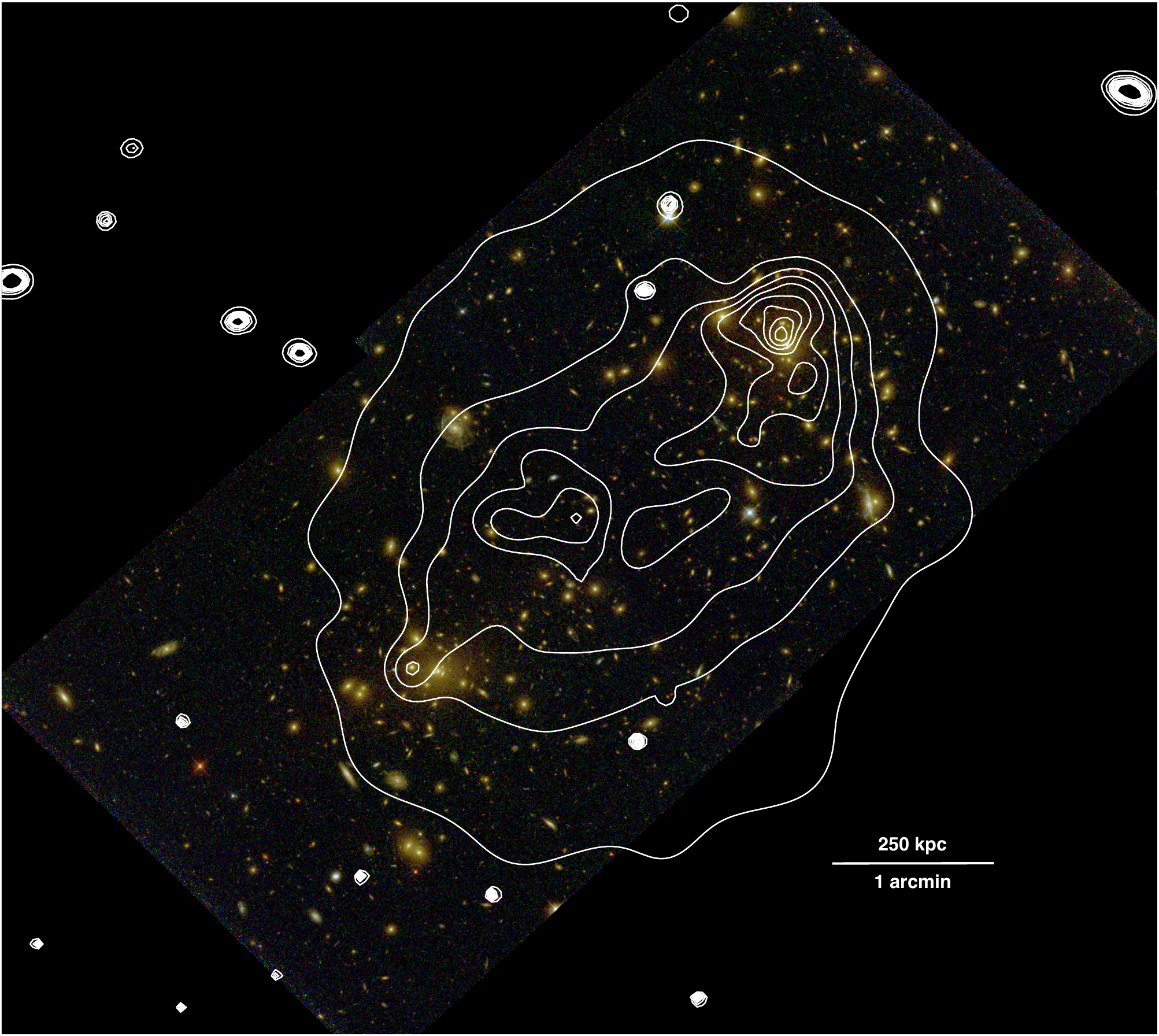}
    \caption{A1758N as seen with HST / ACS (false-colour composite from F435W, F606W, and F814W images collected for GO-12253). Overlaid in white are linearly spaced contours of the adaptively smoothed X-ray surface brightness in the 0.5--7 keV band as observed with Chandra / ACIS-I.}
    \label{fig:a1758-hst}
\end{figure*}

Observations of A1758N with HST's Advanced Camera for Surveys \citep[ACS;][]{ford98} in the F435W, F606W, and F814W filters were performed for GO-12253 (PI: Clowe) in December 2011 for total exposure times of 2536, 2544, and 5000s, respectively. Two observations of the quoted durations were necessary in each filter to cover both components of this merging cluster. Additional shallow coverage at near-infrared wavelengths, again for both of the subclusters, was added by program GO-14096 (PI: Coe) through short exposures (656s, 331s, 431s, and 1056s) with the Wide Field Camera 3 \citep[WFC3;][]{kimble08} in April and June 2016 in the F105W, F125, F140W, and F160W filters, respectively, as part of the RELICS program \citep{coe19}. We use the resulting high-level science products publicly available from the MAST archive; a color image of A1758N based on the cited HST / ACS observations is shown in Fig.~\ref{fig:a1758-hst}.

\subsection{Galaxy spectroscopy}
\label{sec:spec}

Extensive groundbased spectroscopy of galaxies in the A1758N field was performed by \citet{boschin12} and \citet{mo17}. Blissfully unaware of this earlier work, we observed A1758N with the DEIMOS spectrograph on the Keck-II 10m-telescope in poor conditions in July 2018. Three multi-object spectroscopy (MOS) masks were designed to obtain low-resolution spectra of presumed cluster members, potential strong-lensing features, and RPS candidates (see Section~\ref{sec:targets} for details of the target selection). All slits were 1\arcsec\ wide; the instrumental setup combined the 600 l/mm grating (set to a central wavelength of 6300\AA) with the GG455 blocking filter to suppress second-order contributions at $\lambda>9000$\AA.

Although these observations were later found to have duplicated a significant number of measurements already reported in the literature, they corrected one erroneous literature redshift, added 23 new spectra (and redshifts), and allowed us to test for systematic biases as discussed in Section~\ref{sec:analysis}. 

\subsection{X-ray imaging spectroscopy}
\label{sec:xover}
A1758 was observed with the Chandra X-ray Observatory's ACIS-S detector in 2001 (Sequence Number 800152; PI: David) for 58 ks, and with ACIS-I in September and October 2012 (Sequence Number 801177; PI: David)  for a total exposure time of 148 ks. All observations were performed in Very Faint mode. We reprocessed and merged all ACIS-I observations using \textsc{CIAO 4.8}, and then adaptively smoothed the emission in the 0.5--7 keV band to $3\sigma$ significance using the \textsc{Asmooth} algorithm \citep{ebeling06}. The resulting iso-intensity X-ray surface brightness contours are shown in Fig.~\ref{fig:a1758-hst}.

The X-ray properties of both A1758S and A1758N as determined from Chandra and XMM-Newton observations, as well as their significance for the interpretation of the extensive merging activity in this system, are discussed by \citet{david04} and \citet{durret11}.

\subsection{Radio observations}
Diffuse radio emission from A1758N was first detected in the NVSS and WENSS surveys \citep{kempner01}; investigations conducted with the VLA at 1.4\,GHz, the GMRT at 325\,MHz, and LOFAR at 144\,MHz resolve a giant radio halo extending beyond the X-ray emission and across the full extent of Fig.~\ref{fig:a1758-hst} \citep{giovannini09,venturi13,
botteon18}. No radio relics associated with A1758N were detected.

\section{Numerical simulations}
Smoothed-particle hydrodynamics simulations of the active merger A1758N performed by \citet{machado15} found the observed X-ray morphology of the system (Fig.~\ref{fig:a1758-hst}) to be best replicated by a collision of clusters of equal mass\footnote{Estimates for the total mass of A1758N are compiled in \citet{mo17} and range from 1 to 2$\times10^{15}$ M$_\sun$; \citet{machado15} adopt $\sim5\times 10^{14}$ M$_\sun$ for each subcluster.} starting from initial conditions  characterized by a 3 Mpc separation, an impact parameter of 250 kpc, and a relative velocity of 1500 km s$^{-1}$. Requiring the observational constraints to be met when the separation of the two subclusters equals the observed (projected) distance of 750 kpc between the BCGs, \citet{machado15} report that the best match is attained at t=1.7 Gyr, after the first core passage and just before turnaround, when the relative velocity of the subclusters is 380 km s$^{-1}$. 

Since the simulations assumed that the collision proceeds in the plane of the sky, the relative velocity between the two merger components along our line of sight is essentially zero at all times, although \citet{machado15} state that their results remain valid if the collision is viewed at an angle of up to 20 degrees, in which case a relative line-of-sight velocity of up to 130 km s$^{-1}$ is predicted for the two BCGs. As expected for a binary collision of massive clusters, these simulations also predict strong shocks about 600 kpc away from each BCG, propagating outward through the ICM at Mach numbers of at least 6. Consistent findings are reported by \citet{mo17} who refine the simulations by \citet{machado15} while retaining the assumption of a plane-of-the-sky merger.

\section{Data Analysis}
\label{sec:analysis}

\subsection{Photometry}
\label{sec:phot-data}

We use the photometric data provided as a high-level science product by the HST MAST archive at \url{ https://archive.stsci.edu/missions/hlsp/relics/abell1758/catalogs/}, obtained with SExtractor \citep{bertin96} in dual-image mode with the F606W image as the detection band and the settings employed for all data acquired for GO-14096.

\subsection{Spectroscopy}

All spectroscopic data, gathered by us with Keck-II/DEIMOS as described in Section~\ref{sec:spec}, were reduced using a modified version of the DEEP2 pipeline \citep{cooper12,newman13}. In the following we describe the selection of galaxy targets and the determination of cluster membership from the resulting redshifts.

\subsubsection{Target selection}
\label{sec:targets}

\begin{figure}
    \centering
    \hspace*{-5mm}\includegraphics[width=0.5\textwidth]{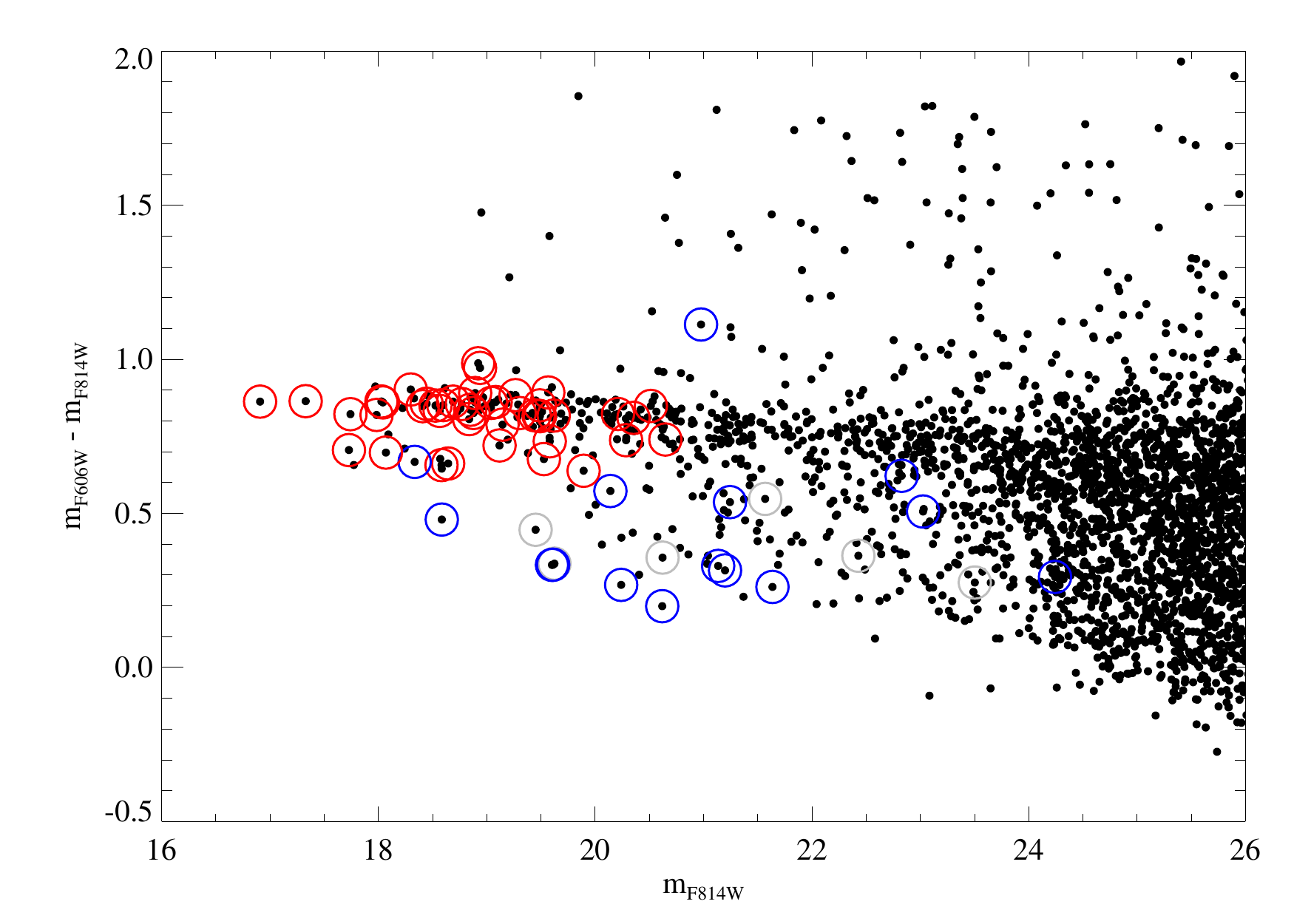}
    \caption{Color-magnitude diagram of all galaxies within the field of Fig.~\ref{fig:a1758-hst}. Blue circles mark galaxies targeted in our DEIMOS observations on the grounds of their disturbed morphology; red circles mark ellipticals observed as likely cluster members; fillers (grey) were mainly selected from the population of bright, blue galaxies.}
    \label{fig:cmd}
\end{figure}

The automatically generated SExtractor source catalog referred to in Section~\ref{sec:phot-data} contains a significant number of  false detections. We excluded 1092 spurious sources at the field edges and then limited the remaining catalog to objects classified as galaxies and meeting the criterion $m_{\rm F814W}<28$. After visual scrutiny of the brightest sources in this list, we removed a further 22 objects with $m_{\rm F814W}<24$ as obvious stars. A color-magnitude diagram of the resulting galaxy sample is shown in Fig.~\ref{fig:cmd}. 
Highlighted are galaxies selected by us for follow-up spectroscopy either because of their disturbed morphology in the HST image (this subset includes both RPS candidates and potential strong-lensing features), or because their location on the cluster red sequence (clearly visible in Fig.~\ref{fig:cmd}) makes them likely cluster members. 

In the selection of RPS candidates we followed the prescription provided by \citet{ebeling14} which considers three primary morphological indicators, namely (1) signs of  unilateral external forces, (2) brightness or color gradients suggesting triggered star formation, and (3) the presence of debris trails. Similar criteria have been used by other authors to select RPS candidates \citep[e.g.,][]{pog16}.

\subsubsection{Redshift measurements and cluster membership}

Redshifts were determined for the 66 galaxies marked in Fig.~\ref{fig:cmd} through cross-correlation with spectral templates and subsequent correction to the heliocentric frame, using an adaptation of the SpecPro package \citep{masters11}.  This data set, presented in full in Appendix A, includes the brightest cluster galaxy (BCG) of either subcluster. From these redshifts, the ROSTAT statistics package \citep{beers90} measures a systemic cluster redshift of $z=0.2775$ for A1758N and a velocity dispersion of $\sigma=1780^{+140}_{-160}$ km s$^{-1}$ from 51 concordant redshifts. 

Spatial cross-correlation of our sample with the positions of 203 galaxies with spectroscopic redshifts listed in \citet{mo17} identified 43 objects as in common. For all but one galaxy\footnote{The exception is the foreground spiral at $(\alpha,\delta)=(13\;32\;56.1,+50\;30\;17)$ (J2000) for which our emission-line redshift of 0.1038 supersedes the erroneous literature value of 0.2764.}, our redshift measurements are in excellent agreement with the literature values ($\langle \Delta z\rangle=-0.0001\pm 0.0005$). 

Our final sample of spectroscopically confirmed cluster members thus comprises 159 galaxies, 51 observed by us and 118 from the compilation by \citet{mo17}.

\subsection{Calibration and extinction correction}
\label{sec:spec-cal}

The DEIMOS spectra were flux calibrated by first dividing the observed spectra by the spectrograph's response function for the grating, blocking filter, and central wavelength used during our observations (see Section~\ref{sec:spec} for details) and then scaling the integrated signal within the ACS/F606W passband such that it matches the observed HST photometry in the same filter, i.e.,
\begin{equation}
f(\lambda) =  \frac{f_{\lambda,{\rm F606W}}}{\int s(\lambda)T_{\rm F606W}(\lambda) d\lambda \; / \; 2235.5{\rm \AA}}\; s(\lambda),\label{eqn:speccal}
\end{equation}
where $f_{\lambda,{\rm F606W}}$ is the flux density derived from the galaxy's isophotal magnitude in the ACS/F606W image, and $f(\lambda)$ and $s(\lambda)$ are the flux-calibrated and the response-corrected observed spectrum, respectively. $T_{\rm F606W}(\lambda)$ and 2235.5\AA\ are the effective throughput and bandwidth (cumulative throughput width, CTW95) of the ACS/F606W filter, respectively. For all spectra that fully contain both the F606W and F814W bandpasses we computed the flux calibration factor of Eqn.~\ref{eqn:speccal} for either filter and found the results to be consistent within 6\%.

We subsequently corrected the spectra for interstellar extinction in the Milky Way following \citet{seaton79} and adopting the E(B--V) reddening coefficient of 0.0118 measured by \citet{schlafly11} in the direction of our cluster target.

Note that this calibration procedure implicitly assumes that the spectrum recorded within each DEIMOS slit is representative of that of the entire galaxy.

\subsection{Emission-Line Ratios, Star-Formation Rates, Stellar Masses}

We measured (net) emission-line fluxes for the thermally excited H$\alpha$ and H$\beta$ lines, as well as for the collisionally excited [\ion{N}{2}] and [\ion{O}{3}] lines, from the calibrated DEIMOS spectra, where possible. The resulting ratios [\ion{N}{2}]$\lambda$6583/H$\alpha$ and [\ion{O}{3}]$\lambda$5007/H$\beta$ will be used as diagnostics in Section~\ref{sec:spec-diagnostics}.

Star-formation rates (SFR) for our RPS candidates were derived from the H$\alpha$ emission-line luminosity, using the scaling relation 
$${\rm SFR}[{\rm M}_{\odot}{\rm yr}^{-1}]=9.9 \times 10^{-42}L({\rm H}\alpha)\; {\rm [erg\,\, s}^{-1}]$$
\citep{ken98}, where $L({\rm H}\alpha)$ is the net luminosity of the H$\alpha$ emission line, corrected for intrinsic extinction as determined from the Balmer decrement following \citet{calzetti2000}.

Finally, we obtained stellar masses with the SED fitting package \textsc{Prospector} \citep{leja17}, using both the HST photometry  (ACS and WFC3 where available) and the calibrated DEIMOS spectrum as observational constraints. Like the DEIMOS spectra (see Section~\ref{sec:spec-cal}), the HST photometry too was corrected for Galactic extinction.

\section{Results}
\label{sec:results}

\subsection{Nature of ram-pressure stripping candidates}
\begin{figure}
    \centering
    \includegraphics[width=0.45\textwidth]{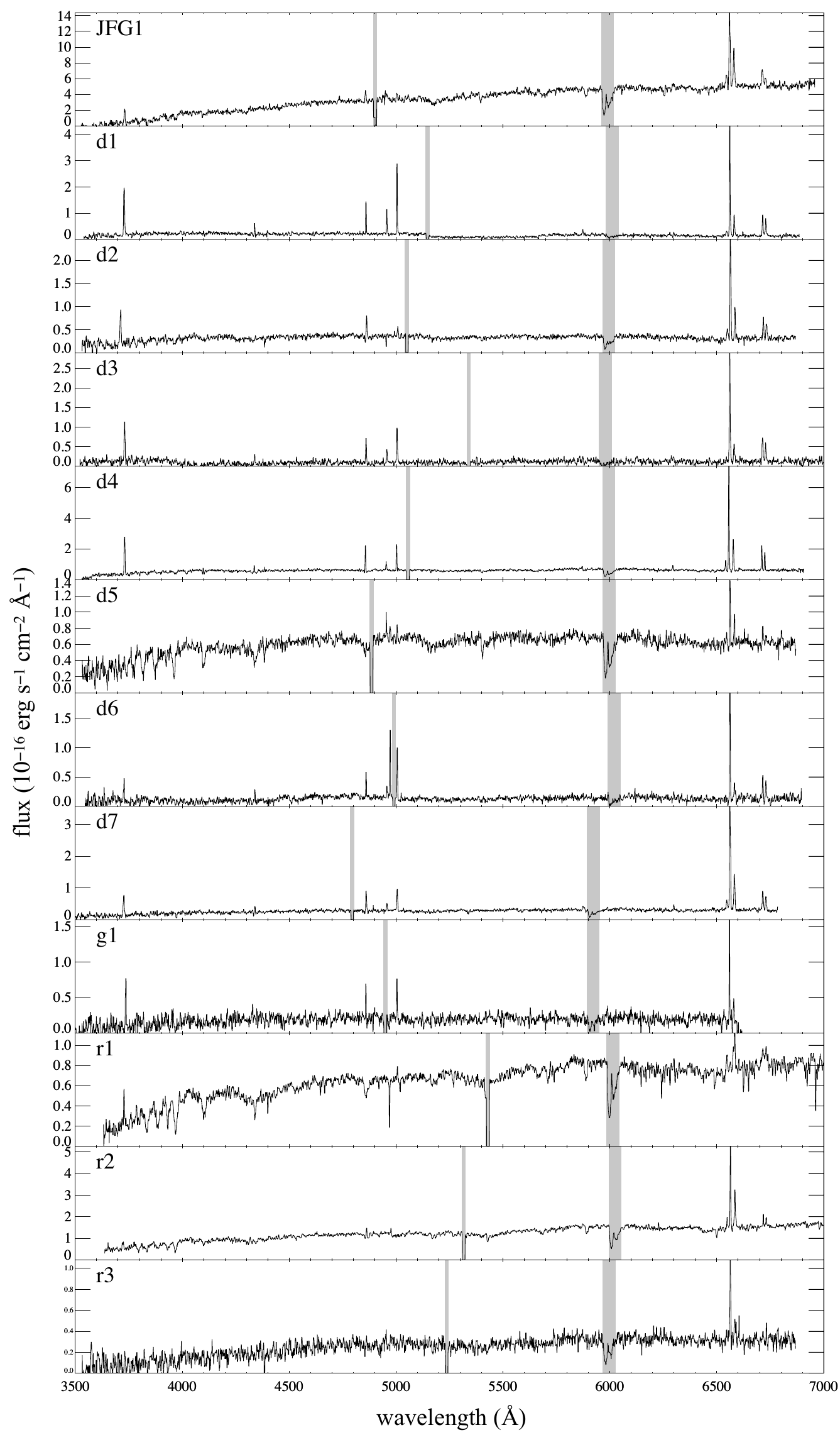}
    \caption{Rest-frame spectra of the galaxies shown in Fig.~\ref{fig:RPS-cands}. The DEIMOS chip gap and absorption at 7600\AA\ from water in the atmosphere are greyed out.}
    \label{fig:1dspec}
\end{figure}

\begin{figure*}
    \centering
    \includegraphics[width=0.32\textwidth]{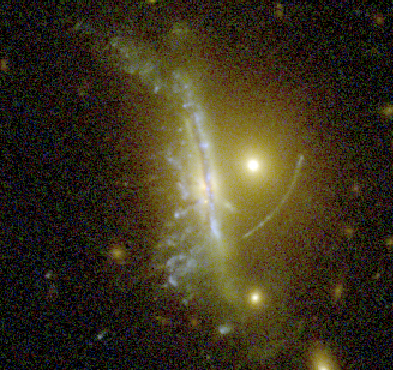}
    \includegraphics[width=0.32\textwidth]{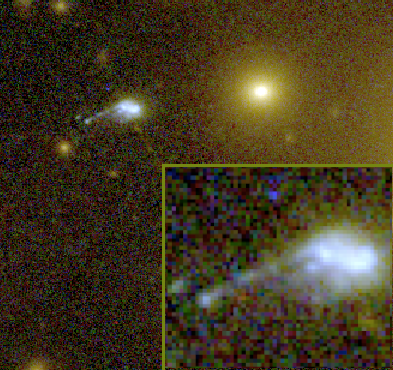}
    \includegraphics[width=0.32\textwidth]{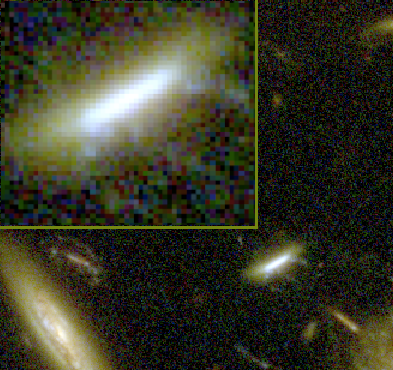}
    \includegraphics[width=0.32\textwidth]{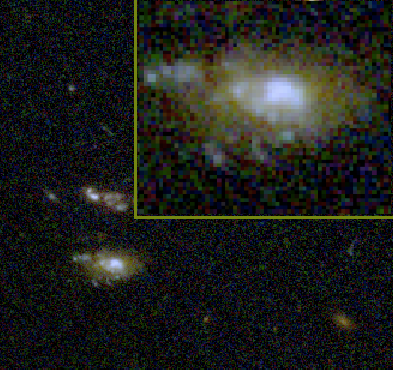}
    \includegraphics[width=0.32\textwidth]{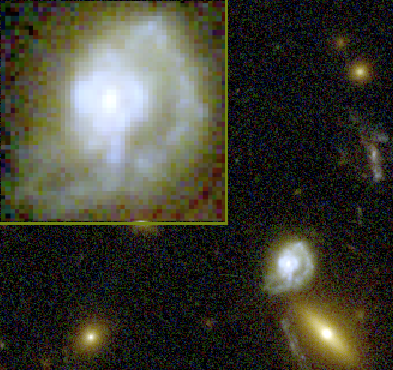}
    \includegraphics[width=0.32\textwidth]{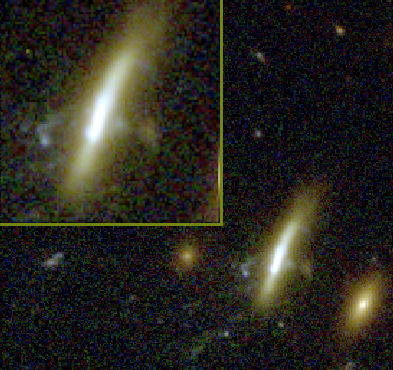}
    \includegraphics[width=0.32\textwidth]{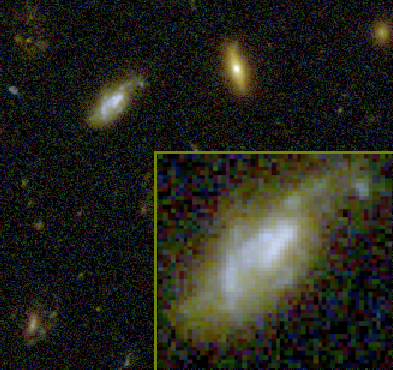}
    \includegraphics[width=0.32\textwidth]{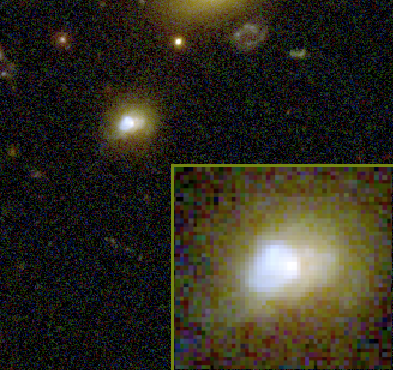}
    \includegraphics[width=0.32\textwidth]{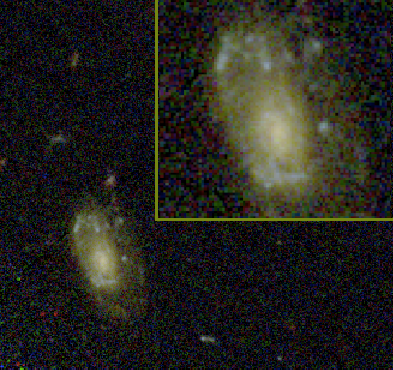}
    \includegraphics[width=0.32\textwidth]{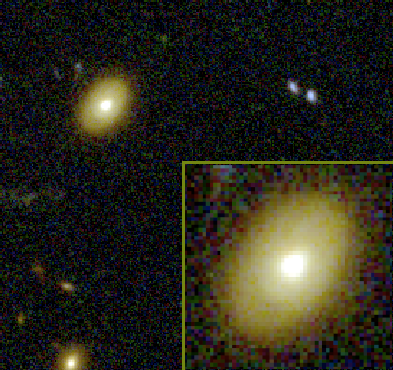}
    \includegraphics[width=0.32\textwidth]{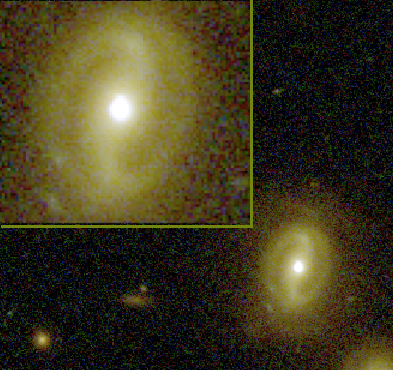}
    \includegraphics[width=0.32\textwidth]{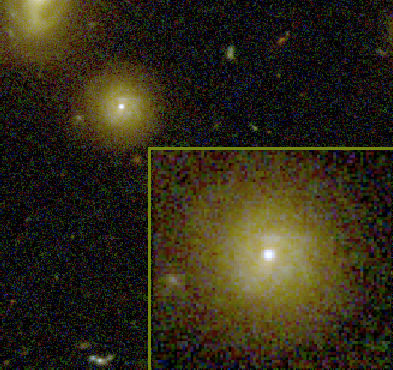}\\
    \vspace{-170mm}
    \hspace*{-50mm}
    \textcolor{white}{\textbf{JFG1}\hspace{0.29\textwidth}\textbf{d1}\hspace{0.3\textwidth}\textbf{d2}}\\[51mm]
    \hspace*{-50mm}
    \textcolor{white}{\textbf{d3}\hspace{0.305\textwidth}\textbf{d4}\hspace{0.305\textwidth}\textbf{d5}}\\[51mm]
    \hspace*{-50mm}
    \textcolor{white}{\textbf{d6}\hspace{0.305\textwidth}\textbf{d7}\hspace{0.305\textwidth}\textbf{g1}}\\[51mm]
    \hspace*{-50mm}
    \textcolor{white}{\textbf{r1}\hspace{0.305\textwidth}\textbf{r2}\hspace{0.305\textwidth}\textbf{r3}}\\[3mm]
    \caption{Close-up view of the galaxies exhibiting H$\alpha$ emission in our DEIMOS spectra. The first eight are JFG1 and seven additional galaxies targeted because their disturbed morphology suggested ongoing ram-pressure stripping (d1 to d7). The final three (r1 to r3) were selected for observation as likely cluster members because of their color, which is consistent with the cluster red sequence. All images span 17 arcsec on a side; where present, insets show a magnified view of the observed galaxy. The resulting DEIMOS spectra are shown in Fig.~\ref{fig:1dspec}.}
    \label{fig:RPS-cands}
\end{figure*}

\begin{figure*}
    \centering
    \includegraphics[width=0.95\textwidth]{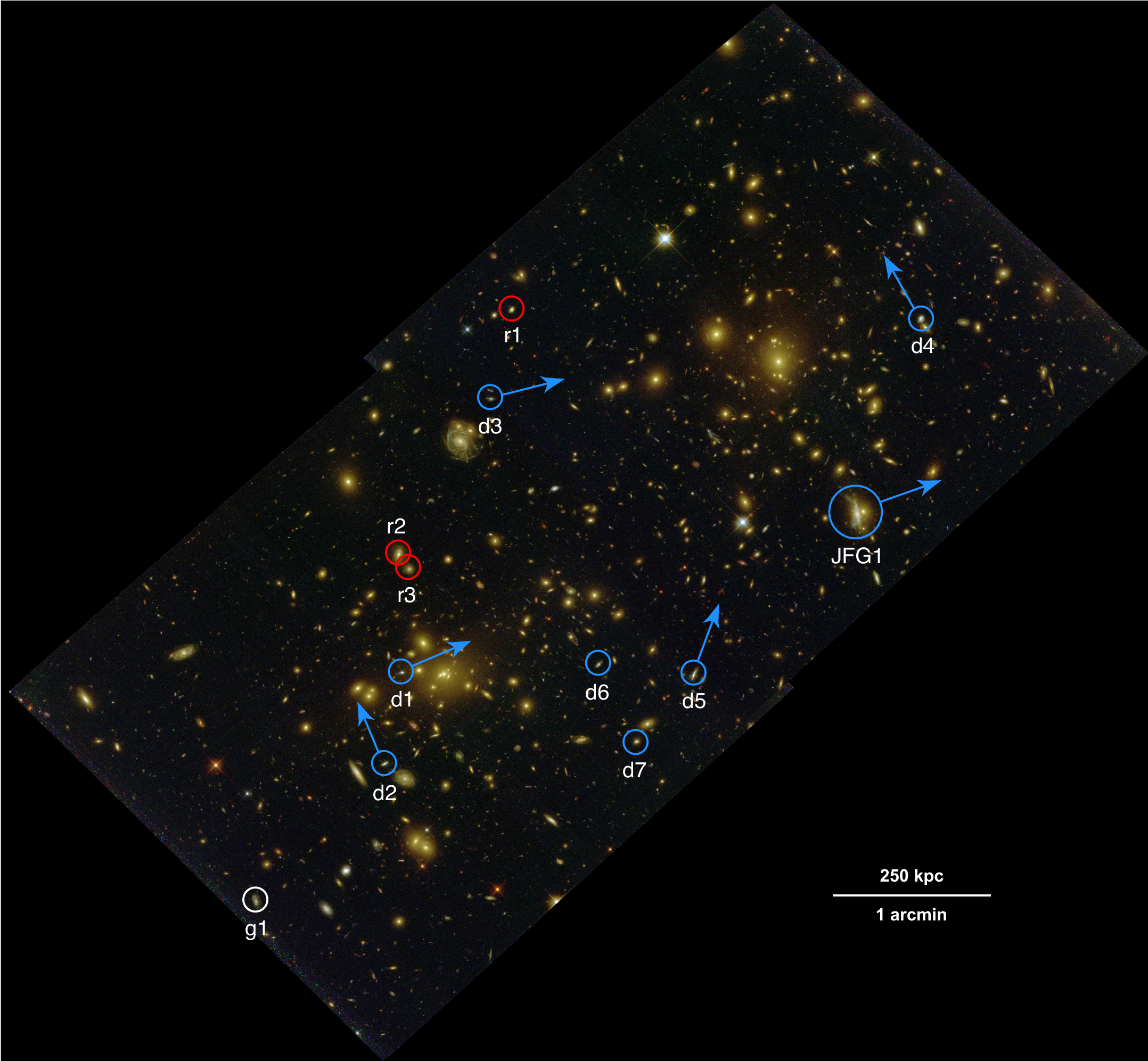}
    \caption{As Fig.~\ref{fig:a1758-hst} but with all galaxies from Fig.~\ref{fig:RPS-cands} marked; color coding as in Fig~\ref{fig:cmd}. Arrows indicate the deduced approximate direction of motion in the plane of the sky for RPS candidates with discernible debris tails.}\label{fig:marked_A1758N}
\end{figure*}

Of the fifteen galaxies targeted by us with DEIMOS because of their disturbed optical morphology, two were found to be foreground galaxies, and five are potentially gravitationally lensed background objects at $z=0.6-1.3$; the remaining eight are cluster members. All of the latter show emission lines, as do four additional cluster members, three of which were targeted because of their red-sequence color. The DEIMOS spectra of the sample of 12 line emitters are shown in Fig.~\ref{fig:1dspec}, their HST images are presented in Fig.~\ref{fig:RPS-cands}, and their location within the cluster is indicated in Fig.~\ref{fig:marked_A1758N}.

\subsection{A1758N: radial-velocity distribution}

\begin{figure}
    \centering
    \hspace*{-6mm}\includegraphics[width=0.53\textwidth]{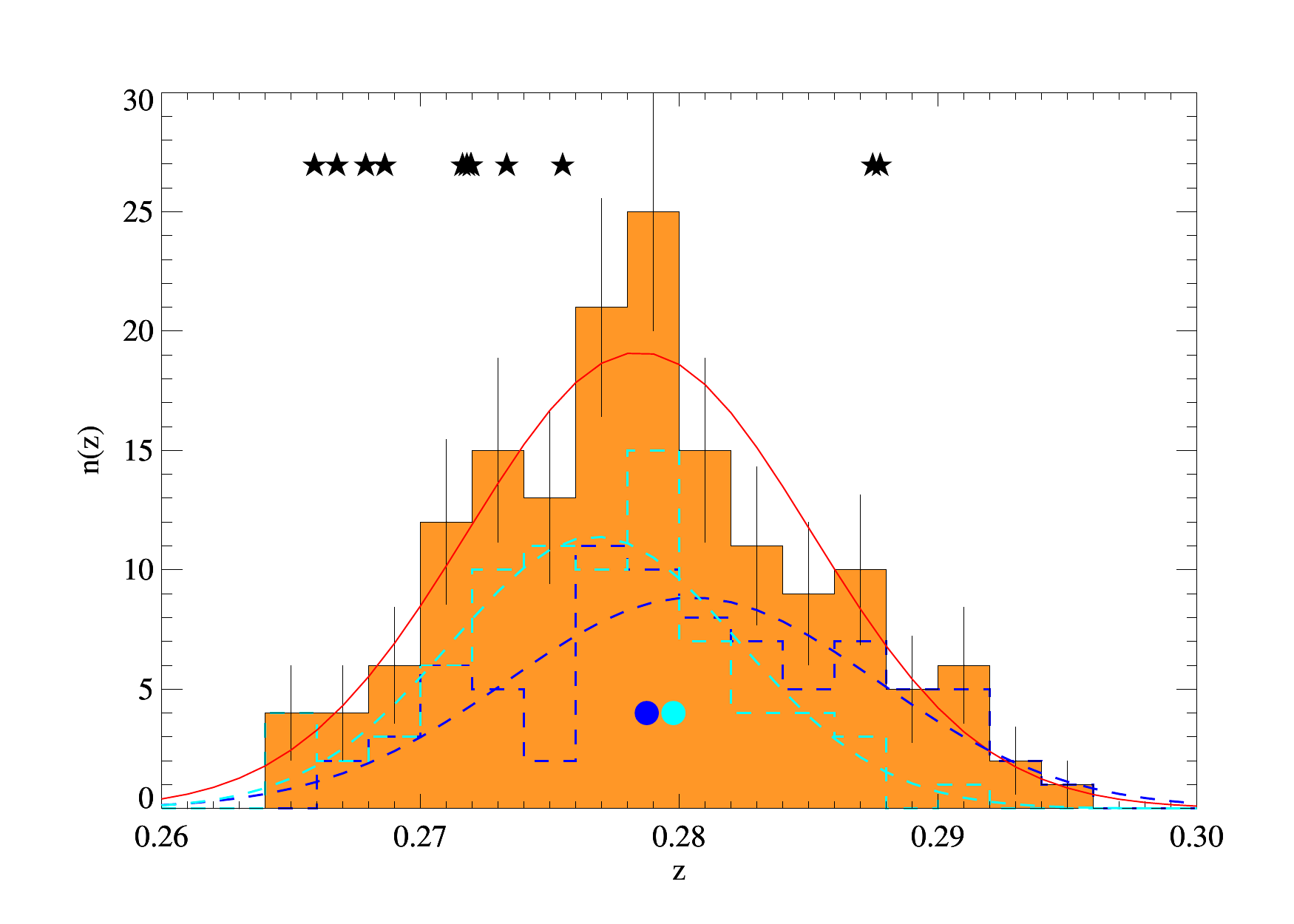}
    \caption{Redshift histogram for 159 spectroscopically confirmed members of A1758N. Although the data are well described by a single Gaussian, overlaid in red, the redshift distribution in the NW half of the field (dashed blue) differs significantly from that in the SE half (dashed cyan). The locations of the respective BCGs, and of all EL galaxies in the DEIMOS data set, are marked at the bottom and top, respectively.}
    \label{fig:zhist}
\end{figure}

The 159 spectroscopically confirmed cluster members in the combined data set yield a cluster redshift of $z=0.2785$ and a velocity dispersion of $\sigma=1553^{+80}_{-100}$ km s$^{-1}$. While the redshift distribution of A1758N's galaxy population (presented in Fig.~\ref{fig:zhist}) shows no sign of bimodality, splitting the galaxy sample, as viewed in projection on the sky, at the midpoint between the two BCGs results in redshift distributions that differ at the 3$\sigma$ confidence level according to a two-sided Kolmogorov-Smirnov test, suggesting a significant, albeit small, line-of-sight velocity of two subclusters relative to each other. Since this simplistic split of the overall galaxy population based on projected position in the sky does not cleanly separate the populations of the two subclusters, regardless of the orientation of the merger axis, the observed velocity difference is merely a qualitative indication of relative motions along our line of sight. Complex velocity substructure was also reported by \citet{boschin12} in a study based on 92 cluster members which, however, did not find a significant difference in the systemic redshifts of the two subclusters of A1758N, consistent with the small radial velocity difference of 180 km s$^{-1}$ between the BCGs (see Fig.~\ref{fig:zhist}). 

\subsection{The emission-line subsample: physical properties}
\label{sec:spec-diagnostics}

\begin{table}
\centering
\begin{tabular}{lcc@{\hspace*{1mm}}@{\hspace*{1mm}}c}
name & log(M$_{\star}$M$_{\odot}^{-1}$) & SFR  & $\Delta v_{\rm rad}$ \\ 
& & (M$_{\odot}$ yr$^{-1}$) & (km s$^{-1}$)\\ \hline
JFG1&     10.87$\pm$0.02 &     47.9 & $\;\;-936$\\ 
d1 &\,\,\,8.47$\pm$0.06 & \,\,\,5.6 & $-1933$\\ 
d2 &\,\,\,9.50$\pm$0.03 & \,\,\,5.9 & $-1355$\\ 
d3 &\,\,\,8.39$\pm$0.04 & \,\,\,6.0 & $\;\;-739$\\ 
d4 &\,\,\,9.83$\pm$0.06 &      15.1 & $-1174$\\ 
d5 &     10.10$\pm$0.05 &      16.6 & $-1410$\\ 
d6 &\,\,\,8.88$\pm$0.02 & \,\,\,4.8 & $-2257$\\ 
d7 &\,\,\,9.78$\pm$0.01 &      13.8 & $\;\;\,1320$\\ 
g1 &\,\,\,9.47$\pm$0.01 & \,\,\,1.6 & $\;\;\,1370$\\ 
r1 &     10.46$\pm$0.10 &$>1.1^\star$ & $-1885$\\ 
r2 &     10.62$\pm$0.12 &      12.3 & $-2409$\\ 
r3 &     10.02$\pm$0.06 & \,\,\,6.0 & $-1381$ 
\end{tabular}
\caption{Stellar mass, star formation rate, and radial velocity relative to the (in projection) closest BCG of all galaxies in our DEIMOS sample that show H$\alpha$ emission (see Fig.~\ref{fig:RPS-cands}). Galaxy coordinates and redshifts can be found in Appendix~A. We assign a name to the only indisputable case of RPS and denote the provenance of all other galaxies in our line-emitter sample by a leading ``d" (for disturbed), ``r'' (for red sequence), and ``g'' (for other galaxies).\\
\textit{$^{\star}$ lower limit since negligible H$\beta$ emission prevents computation of the Balmer decrement and hence correction for dust extinction}\label{tab:galz}}
\end{table}

\label{sec:sSFR-calc}

\begin{figure}
    \centering
    \hspace*{-3mm}\includegraphics[width=0.5\textwidth]{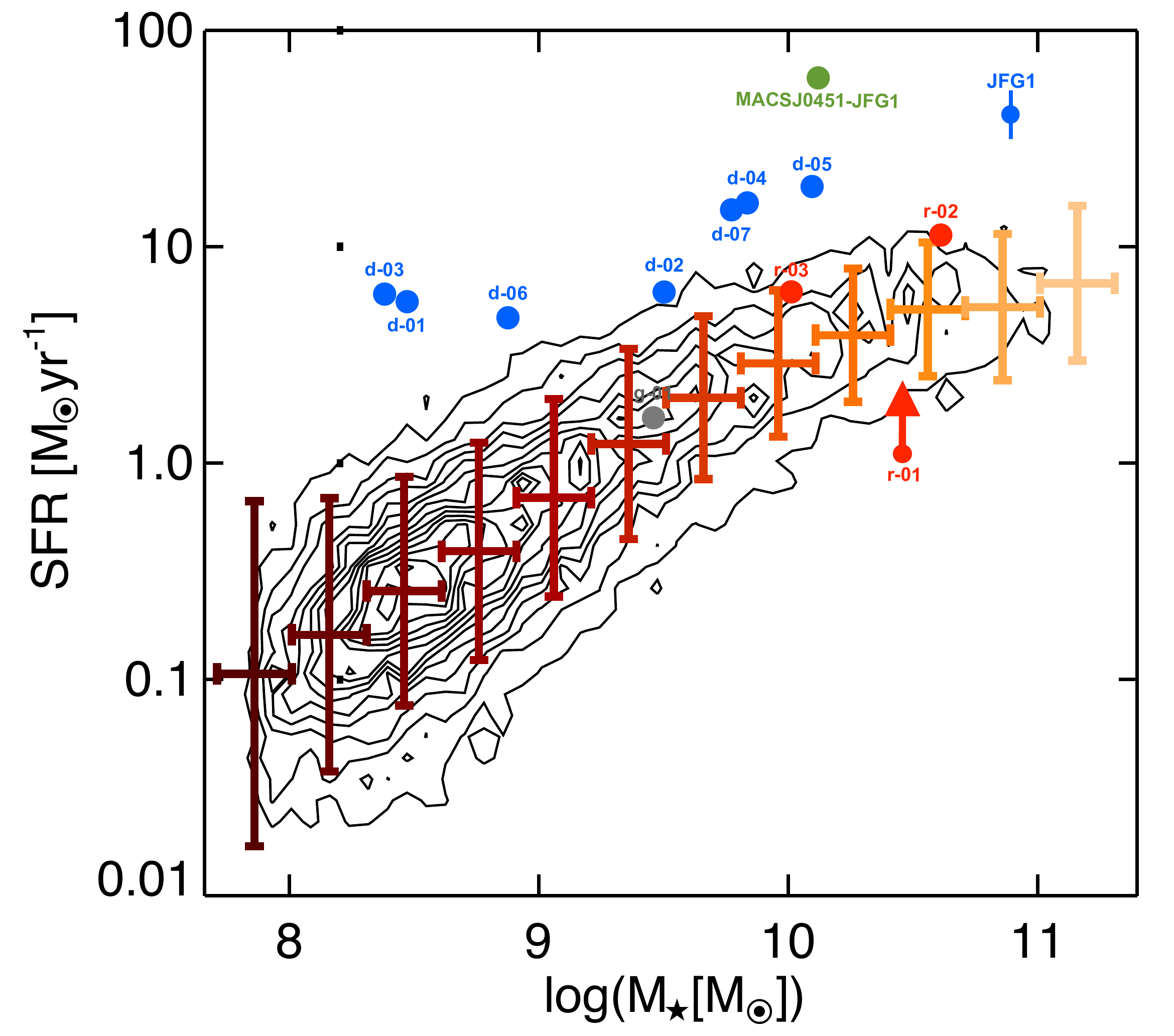}
    \caption{The loci of the 12 galaxies shown in Fig.~\ref{fig:RPS-cands} overplotted on the SFR vs $M_\ast$ relation (figure adapted from \citealt{lee15}). Also shown for reference is MACSJ0451-JFG1 (in green), a textbook example of ram-pressure stripping at $z=0.43$ \citep[][Ebeling et al.\ in prep]{ebeling14}.}
    \label{fig:sSFR}
\end{figure}

A clear physical division within the sample of emission-line galaxies becomes apparent when we examine star formation as a function of stellar mass. As shown in Fig.~\ref{fig:sSFR}, all of the eight galaxies targeted because of morphological signs of RPS in the HST images (see Fig.~\ref{fig:RPS-cands}) lie well above even the outermost confines of the so-called main sequence \citep[e.g.,][]{lee15}, most of them dramatically so. By contrast, the three emission-line galaxies appearing morphologically undisturbed and featuring colors consistent with the cluster red sequence show no elevated star formation\footnote{We note that our determination of the star-formation rate of r01 remains a lower limit because of our inability to obtain a credible estimate of the H$\beta$ flux and hence of the Balmer decrement.} (and neither does g1, observed as a ``filler" on our DEIMOS masks). 

\begin{figure}
    \centering
    \hspace*{-3mm}\includegraphics[width=0.5\textwidth]{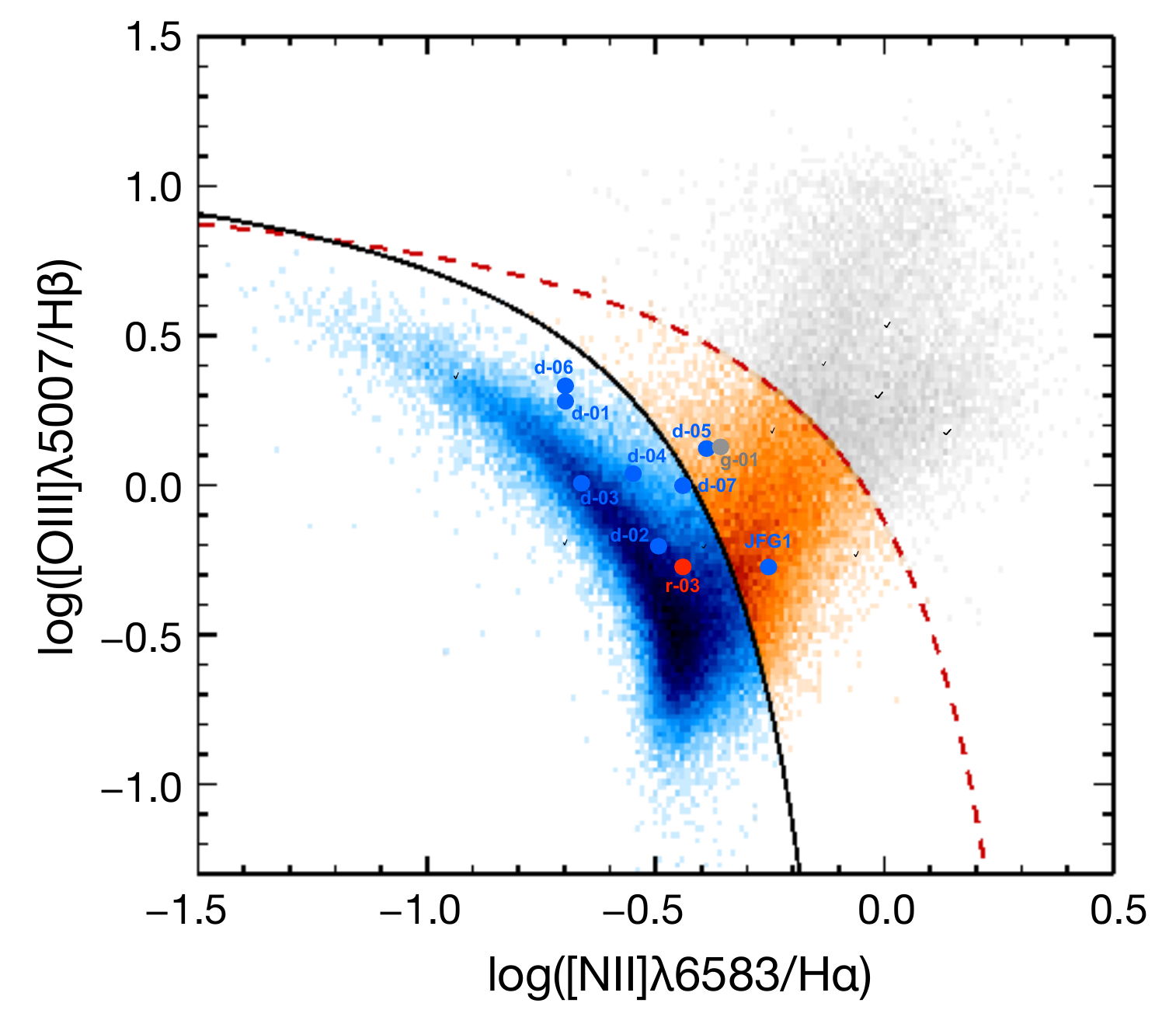}
    \caption{Location of the RPS candidates shown in Fig~\ref{fig:RPS-cands} in the BPT diagram (adapted from \citealt{lara-lopez10}). The solid line shows the empirical division between star-forming and composite galaxies from \citet{kauffmann03}; the dashed line represents the starburst limit from \citet{kewley01}. }
    \label{fig:BPT}
\end{figure}

Used as diagnostics in the BPT diagram (Fig.~\ref{fig:BPT}), the emission line ratios [\ion{N}{2}]$\lambda 6583$/H$\alpha$ and [\ion{O}{3}]$\lambda 5007$/H$\beta$ place all of the galaxies of our emission-line subsample (see Figs.~\ref{fig:1dspec}, \ref{fig:RPS-cands}, \ref{fig:marked_A1758N}) within the star-formation regime \citep{kewley01}. Dividing this regime further, following \citet{kauffmann03}, we find the two most massive galaxies targeted because of their disturbed morphology (JFG1, discussed in detail by \citealt{kalita19}, and d05, see Table~\ref{tab:galz} and Fig.~\ref{fig:sSFR}) to lie in the ``composite" region occupied by galaxies that show both star formation and additional nuclear activity. Note that not all galaxies from Table~\ref{tab:galz} and Fig.~\ref{fig:sSFR} are shown in Fig.~\ref{fig:sSFR} since some lack measurable intensities in the required emission lines.

\section{Discussion}

In the previous section we summarized the properties of the eight RPS candidates in A1758N selected by us for spectroscopic follow-up. Remarkably, they share characteristics that not only strongly support the notion that these systems are indeed undergoing ram-pressure stripping; their dynamical properties also suggest a bulk flow of galaxies that, if real, casts doubt on the merger geometry and dynamics widely adopted for A1758N in the past. In this section we interpret the findings presented in Section~\ref{sec:results}.

\subsection{Triggered star formation and nuclear activity}

Fig.~\ref{fig:sSFR} and Fig.~\ref{fig:BPT} demonstrate that all eight RPS candidates observed by us are undergoing a period of intense star formation. In addition, the location of the most massive galaxies among our targets in the composite region of the BPT diagram (Fig.~\ref{fig:BPT}) is consistent with modest nuclear activity, in agreement with interplay and possibly a causal link between RPS and AGN as pointed out by \citet{pog17} and \citet{george19}. A caveat regarding classifications based on the BPT diagram is in order though: our RPS candidate d07 (just outside the composite region in Fig.~\ref{fig:BPT}) is found to be an X-ray bright point source in the archival Chandra data (Fig.~\ref{fig:a1758-hst}), which unambiguously identifies this galaxy as an AGN. Nonetheless, the combination of these systems' disturbed morphology and extreme star-formation rates strongly supports our initial RPS hypothesis for most of our targets, in particular since galaxy mergers, the only other plausible physical process that could explain the observational evidence, are highly improbable due to the very low cross section for galaxy collisions at the high peculiar velocities encountered in massive clusters.

\subsection{Direction of motion}
\label{sec:motion}

While the spatial distribution (as projected onto the plane of the sky) of our eight RPS galaxies does not exhibit any obvious pattern, their projected direction of motion, as deduced from their debris trails (where discernible), shows a clear tendency to point toward the NW, as indicated in Fig.~\ref{fig:marked_A1758N} (six out of eight\footnote{Note that, while our estimate of the direction of motion as indicated in Fig.~\ref{fig:marked_A1758N} is somewhat subjective, we here only consider the velocity component along the line connecting the two BCGs of A1758N; i.e., our classification is a binary one: toward the NW or toward the SE.}, a ratio that has a probability of 10.9\% of occurring by chance in an isotropic distribution). Moreover, all of them also feature high radial velocities relative to both BCGs, and for most of them (seven out of eight) these peculiar velocities are negative (Table~\ref{tab:galz} and Fig.~\ref{fig:zhist}), indicative of a rapid motion toward us. The probability of this distribution occurring by chance in a radial velocity distribution centered on the systemic cluster redshift is 3.1\%. Since the radial and transverse velocity components are statistically independent, the combined probability of the observed distribution of velocity vectors being a coincidence is $3\times 10^{-3}$. While this number (which formally corresponds to 2.9$\sigma$ significance) ought to be taken with a grain of salt, primarily because of the subjective interpretation of the orientation of the potential debris trails, the observed distribution of velocities for the sample of emission-line galaxies is clearly far from isotropic. The implications of this result are discussed in Section~\ref{sec:mergerimpact}. 
We further note that, since at least four of the RPS candidates (JFG1, d1, d2, and d3) feature clearly visible and credible debris trails, the significant (but unknown) velocity component in the plane of the sky of these systems makes their total three-dimensional velocities in the cluster restframe even higher than the observed radial velocities of 700 to 2000 km s$^{-1}$. Moreover, the observed large radial peculiar velocities imply that the debris trails of these systems are likely to be much longer in three dimensions than the observed, projected lengths of 10 to 30 kpc (see Fig.~\ref{fig:RPS-cands}), rendering these galaxies potentially extreme cases of RPS akin to ESO 137-001 in the nearby Universe \citep{sun07,fumagalli14}. Although unambiguous evidence of ongoing RPS events is lacking (or less compelling) for the remaining four candidates, the location of all eight galaxies in the SFR vs $M_\ast$ plane (Fig.~\ref{fig:sSFR}) is indicative of vigorous starbursts that are consistent with, but greatly exceed, SFR enhancements found for RPS events in less massive clusters \citep{vulcani18}.

\subsection{The impact of cluster mergers}
\label{sec:mergerimpact}
The observed strong bias in favor of negative peculiar velocities highlighted in Section~\ref{sec:motion} (see also Table~\ref{tab:galz}) is noteworthy, since it effectively rules out two possible scenarios regarding the dynamics and origin of the RPS population in A1758N, both of which should result in a largely symmetric velocity distribution. They are: (a) isotropic infall of late-type galaxies from the surrounding field, and (b) stripping of gas-rich cluster members caused by the high velocities created by a merger proceeding in, or very close to, the plane of the sky (the geometry assumed by all previous studies of this system, including the simulations by \citealt{machado15} and \citealt{mo17}).

Although the lopsided distribution and high amplitudes of radial velocities of our RPS candidates heavily disfavor the specific merger geometry adopted historically for A1758N, the observational evidence nonetheless strongly supports a causal link between RPS and merger events in clusters. Previous studies found two opposing effects of such a link: \citet{stroe15,stroe17} and \citet{ruggiero19} report evidence of  RPS triggered by merger-induced shocks, while other studies \citep{pranger14,deshev17} find the fraction of star-forming galaxies in  mergers reduced compared to non-merging clusters, possibly after a preceding short starburst phase. Although seemingly in conflict with each other at face value, these findings might be reconciled as part of a bigger picture in which RPS events in mergers first trigger an initial burst of star formation, and then reduce or completely quench star formation as the supply of atomic and molecular gas is either exhausted or removed from the affected galaxies. The prominence and duration of either phase are, however, likely to depend on details of the merging systems, such as mass, time since first core passage, and collision geometry. In the following section we describe how a combination of both of these effects can explain the phase-space distribution of RPS candidates in the A1758N merger. 

\subsection{RPS as a diagnostic tool: A history of A1758N?}

\begin{figure*}[t]
    \centering
    \hspace*{0mm}\includegraphics[width=0.98\textwidth]{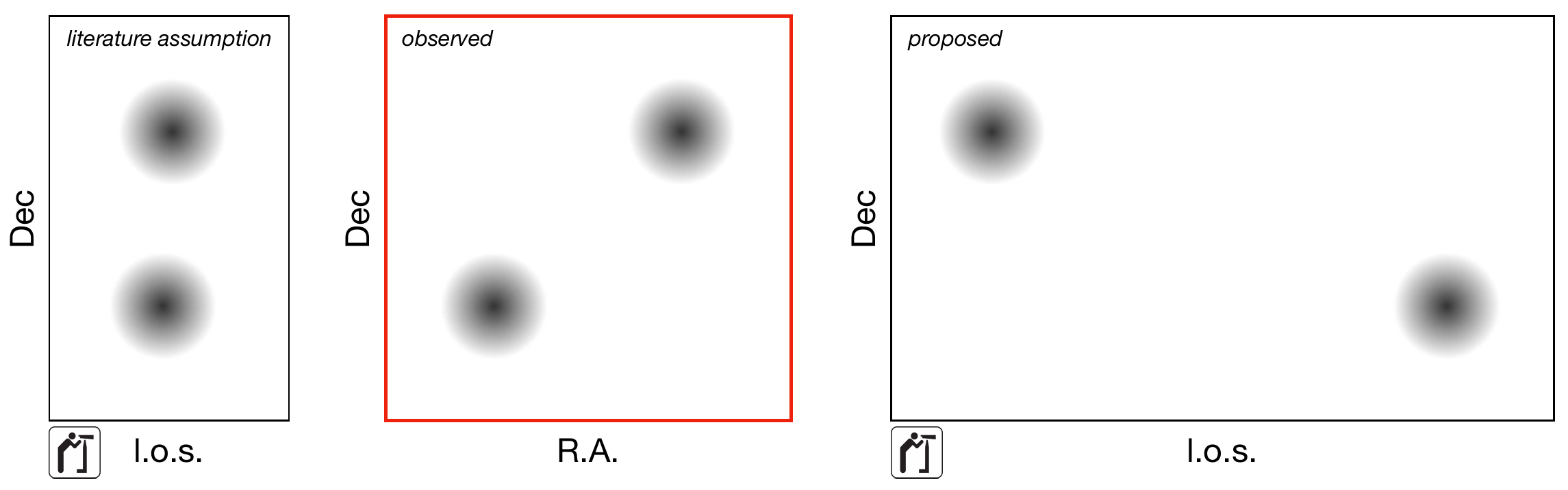}
    \caption{Center: schematic sketch of the observed locations of the two subclusters of the A1758N merger as seen in projection on the sky. The panels on either side show a rotated view that reveals the two components' separation along our line of sight (l.o.s.); the geometry shown in the left panel corresponds to a merger in the plane of the sky, as assumed in the literature; the one on the right illustrates the scenario proposed by use here, in which the merger (observed near turnaround) proceeds along an axis that is substantially inclined with respect to the plane of the sky. }
    \label{fig:sketch}
\end{figure*}

For the majority of clusters, it is primarily the infall of gas-rich galaxies from the surroundings that gives rise to RPS  events. In the absence of pronounced filaments, the resulting velocity distribution is expected to be approximately isotropic, an expectation that clearly is not met by A1758N. Since galaxy infall from the field is unavoidable for a cluster as massive as A1758N, the resulting isotropic RPS distribution is either not well enough sampled to contribute discernibly to our small sample\footnote{Since observational limitations, such as magnitude limit and angular resolution, pose challenges to the identification of RPS events, study of a single cluster may not yield a large enough sample to robustly probe the velocity distribution.}, or the A1758N collision is sufficiently advanced for merger-driven shocks to have quenched star formation in the pre-merger RPS population \citep{deshev17}. The latter possibility finds additional support in the findings by \citet{haines09} who report the presence of a significant and distinct population of passive spiral galaxies in their infrared study of this system.

While the absence of an isotropic velocity distribution in the RPS population of A1758N thus may (but need not)  be a direct consequence of the system's merger history, the cause of the specific, observed directional velocity bias is almost certainly tied to the three-dimensional geometry and environment of the ongoing collision, more specifically to the only preferred direction in this system: the merger axis. Since late-type (gas-rich) galaxies are rare in massive clusters, the population undergoing RPS in A1758N must originate from a much less dense environment, suggesting infall along a filament (rather than isotropically from the general field). However, as the orientation of large-scale filaments determines and indicates the direction along which clusters accrete matter at the vertices of the cosmic web, the direction of the bulk flow of the RPS population of A1758N also marks the most probable orientation of the merger axis. It follows that the latter must be strongly inclined with respect to the plane of the sky, with the feeding filament extending behind and to the SE of the cluster center, as illustrated in Fig.~\ref{fig:sketch}. 

In this scenario, A1758N is not merging in the plane of the sky (as assumed so far in the literature based on the argument that the absence of a significant difference in redshift between the subclusters is evidence of them moving perpendicular to our line of sight) but, in a geometry not unlike that of MACSJ0553.4$-$3342 \citep{ebeling17} along a greatly inclined axis, rendering the observed projected distance between the BCGs of 750 kpc a severe underestimate of their true three-dimensional separation (Fig.~\ref{fig:sketch}). In this scenario, the lack of a pronounced difference in radial velocity between the subclusters and their BCG represents strong evidence of the merger being observed near turnaround.

\section{Summary}

Following the discovery of A1758N\_JFG1 \citep{kalita19}, we identified over a dozen other promising RPS candidates in HST/ACS images of the massive cluster merger A1758N, seven of which were spectroscopically confirmed as cluster members in our DEIMOS observation. Spanning almost three orders of magnitude in stellar mass, all galaxies in the resulting sample of eight feature very high star-formation rates well outside the range observed in regular late-type galaxies. Although only half of these RPS candidates exhibit credible debris trails (an unambiguous sign of ram-pressure stripping), the observed starbursts are unlikely to be the result of minor mergers even for the remaining four, considering the high relative velocities of galaxies (and thus low cross-section for collisions) in this massive cluster merger. Our findings thus lend strong observational support to the notion that RPS, at least in the extreme environment provided by a collision of massive clusters, does not merely displace pre-existing star-forming regions, but in fact triggers powerful starbursts. 

Our sample of galaxies undergoing RPS in A1758N exhibits a highly anisotropic velocity distribution, both along the line of sight and in projection of the sky, suggesting a bulk motion of galaxies relative to the intra-cluster medium along an axis that is strongly inclined with respect to the plane of the sky. The physically most plausible causes for such a pronounced directional bias are galaxy infall along an attached filament or (less compelling, since acting only for a short time) a shock front traveling along the merger axis. Either scenario (or a combination of both) implies that, contrary to all assumptions made to date in the literature, the merger axis of A1758N does not lie in (or close to) the plane of the sky, and that the observed small difference in radial velocity between the subclusters is instead indicative of a collision along a highly inclined merger axis being viewed near turnaround.

Extending our findings and conclusions to cluster mergers in general, we advance the hypothesis that measurements of the peculiar velocities of RPS candidates could be used to constrain the three-dimensional orientation of the merger axis in cluster collisions. Systematic study of the velocities of morphologically disturbed galaxies in other cluster mergers will allow us to test the validity of our hypothesis.

\acknowledgments
BK gratefully acknowledges financial support from the Sheila Watumull Astronomy Fund during his time as a visiting researcher at IfA. We thank Conor McPartland for advice regarding the use of \textsc{Prospector}.

\appendix

\section{Radial velocities}
We list in Table~\ref{tab:allgals} the coordinates and redshifts of all galaxies observed by us with Keck-II/DEIMOS. Details of the instrumental setup are summarized in Section~\ref{sec:spec}.

\begin{table*}
\centering
\begin{tabular}{lcccc|lcccc}
name & R.A.\ & Declination &  $z$ & d$z$ & name & R.A.\ & Declination &  $z$ & d$z$ \\ 
 & \multicolumn{2}{c}{(J2000)} & & & & \multicolumn{2}{c}{(J2000)} & & \\ \hline
 & 13:32:27.373 &+50:34:03.48 & 0.3787 & 0.0002 &        d7 & 13:32:44.222 &+50:31:07.66 & 0.2875 & 0.0001 \\       
 & 13:32:27.900 &+50:34:06.00 & 0.3886 & 0.0003 &         & 13:32:44.934 &+50:31:57.23 & 0.2806 & 0.0005 \\         
 & 13:32:29.199 &+50:34:10.10 & 0.2785 & 0.0002 &         & 13:32:44.956 &+50:34:05.82 & 0.2886 & 0.0002 \\         
 & 13:32:32.075 &+50:32:52.75 & 0.3901 & 0.0002 &         & 13:32:45.327 &+50:33:24.70 & 0.2862 & 0.0002 \\         
d4 & 13:32:32.582 &+50:33:52.49 & 0.2720 & 0.0002 &      d6 & 13:32:45.744 &+50:31:37.79 & 0.2668 & 0.0002 \\       
 & 13:32:32.624 &+50:34:27.88 & 0.1771 & 0.0001 &         & 13:32:45.913 &+50:32:04.74 & 0.2672 & 0.0002 \\         
 & 13:32:34.350 &+50:32:11.28 & 0.2676 & 0.0008 &         & 13:32:46.977 &+50:32:02.01 & 0.2808 & 0.0003 \\         
 & 13:32:34.468 &+50:33:18.54 & 0.2830 & 0.0002 &         & 13:32:48.696 &+50:31:21.69 & 0.2776 & 0.0002 \\         
 & 13:32:34.932 &+50:32:37.28 & 0.2800 & 0.0001 &         & 13:32:48.890 &+50:34:09.58 & 0.2746 & 0.0002 \\         
JFG1 & 13:32:35.169 &+50:32:36.43 & 0.2733 & 0.0003 &    r1 & 13:32:49.296 &+50:33:56.02 & 0.2679 & 0.0003 \\       
 & 13:32:35.357 &+50:32:52.63 & 0.6188 & 0.0004 &        d3 & 13:32:50.138 &+50:33:21.22 & 0.2755 & 0.0002 \\       
 & 13:32:35.476 &+50:34:51.52 & 0.6330 & 0.0004 &         & 13:32:50.992 &+50:33:08.86 & 0.2817 & 0.0003 \\         
 & 13:32:35.720 &+50:33:28.64 & 1.0349 & 0.0001 &         & 13:32:51.958 &+50:31:47.65 & 0.2800 & 0.0008 \\         
 & 13:32:36.486 &+50:32:34.79 & 0.2711 & 0.0004 &    BCG-E  & 13:32:52.065 &+50:31:33.98 & 0.2798 & 0.0001 \\         
 & 13:32:36.619 &+50:32:06.65 & 0.2693 & 0.0010 &         & 13:32:52.815 &+50:30:26.44 & 0.2823 & 0.0002 \\         
 & 13:32:36.700 &+50:33:59.24 & 0.3280 & 0.0001 &         & 13:32:52.901 &+50:31:46.05 & 0.2659 & 0.0006 \\         
 & 13:32:37.560 &+50:33:05.77 & 0.2744 & 0.0002 &         & 13:32:53.293 &+50:31:13.77 & 1.0809 & 0.0001 \\         
 & 13:32:38.305 &+50:31:30.22 & 0.2886 & 0.0006 &        r3 & 13:32:53.488 &+50:32:14.98 & 0.2718 & 0.0003 \\       
 BCG-W & 13:32:38.395 &+50:33:35.72 & 0.2787 & 0.0001 &         & 13:32:53.652 &+50:30:53.32 & 0.3299 & 0.0003 \\         
 & 13:32:38.448 &+50:31:41.49 & 0.2838 & 0.0003 &        d1 & 13:32:53.785 &+50:31:34.65 & 0.2686 & 0.0001 \\       
 & 13:32:38.550 &+50:33:43.93 & 0.2792 & 0.0005 &         & 13:32:53.900 &+50:29:57.62 & 0.2738 & 0.0003 \\         
 & 13:32:39.413 &+50:34:45.08 & 0.2776 & 0.0002 &        r2 & 13:32:53.918 &50:32:20.56 & 0.2659 & 0.0003 \\        
 & 13:32:39.526 &+50:34:32.00 & 0.2936 & 0.0002 &         & 13:32:54.387 &+50:33:34.25 & 0.2723 & 0.0004 \\         
 & 13:32:39.553 &+50:34:00.20 & 0.2842 & 0.0002 &        d2 & 13:32:54.475 &50:30:59.13 & 0.2719 & 0.0002 \\        
 & 13:32:39.759 &+50:32:41.07 & 0.2789 & 0.0002 &         & 13:32:54.785 &+50:30:27.78 & 0.1753 & 0.0001 \\         
 & 13:32:40.481 &+50:35:39.69 & 0.2727 & 0.0001 &         & 13:32:55.064 &+50:32:04.84 & 0.2751 & 0.0004 \\         
 & 13:32:40.640 &+50:34:51.46 & 1.2823 & 0.0001 &         & 13:32:55.122 &+50:31:25.43 & 0.2838 & 0.0001 \\         
 & 13:32:40.939 &+50:33:46.29 & 0.2788 & 0.0001 &         & 13:32:55.975 &+50:32:49.03 & 0.2646 & 0.0001 \\         
d5 & 13:32:41.868 &+50:31:33.58 & 0.2716 & 0.0001 &       & 13:32:56.058 &+50:30:17.36 & 0.1038 & 0.0000 \\         
 & 13:32:42.021 &+50:34:34.77 & 0.2923 & 0.0004 &         & 13:32:56.875 &+50:30:02.76 & 0.1504 & 0.0002 \\         
 & 13:32:43.364 &+50:33:05.26 & 1.2472 & 0.0001 &         & 13:32:57.024 &+50:32:13.14 & 0.2851 & 0.0003 \\         
 & 13:32:43.415 &+50:33:28.68 & 0.2853 & 0.0003 &         & 13:32:57.701 &+50:31:13.23 & 0.2804 & 0.0002 \\         
 & 13:32:43.764 &+50:31:14.63 & 0.2778 & 0.0002 &       g1 & 13:32:59.725 &+50:30:05.24 & 0.2878 & 0.0006 \\ \hline
\end{tabular}
\caption{Positions and redshifts (with associated uncertainties) of all galaxies observed by us with Keck-II/DEIMOS. The 12 emission-line galaxies are labelled. \label{tab:allgals}}
\end{table*}

\end{document}